\documentclass[conference]{IEEEtran}

\IEEEoverridecommandlockouts

\makeatletter
\newcommand{\biggg}[1]{{\hbox{$\left#1\vbox to 20.5pt{}\right.\n@space$}}}
\newcommand{\Biggg}[1]{{\hbox{$\left#1\vbox to 23.5pt{}\right.\n@space$}}}
\newcommand{\bigggg}[1]{{\hbox{$\left#1\vbox to 26.5pt{}\right.\n@space$}}}
\newcommand{\Bigggg}[1]{{\hbox{$\left#1\vbox to 29.5pt{}\right.\n@space$}}}
\newcommand{\biggggg}[1]{{\hbox{$\left#1\vbox to 32.5pt{}\right.\n@space$}}}
\newcommand{\Biggggg}[1]{{\hbox{$\left#1\vbox to 35.5pt{}\right.\n@space$}}}
\newcommand{\bigggggg}[1]{{\hbox{$\left#1\vbox to 38.5pt{}\right.\n@space$}}}
\newcommand{\Bigggggg}[1]{{\hbox{$\left#1\vbox to 41.5pt{}\right.\n@space$}}}
\makeatother

\usepackage{bm,cite,algorithm,algorithmic,float,amsmath,amssymb}
\usepackage[dvips]{graphicx}
\usepackage{subfigure,amsfonts}
\usepackage{mdwmath}
\usepackage{mdwtab}
\usepackage{xcolor}
\usepackage{balance}
\floatname{algorithm}{Algorithm}

\allowdisplaybreaks
\begin{document}


\title{Rate Splitting Multiple Access Aided Mobile Edge Computing in Cognitive Radio Networks}

\author{
\IEEEauthorblockN{
Hongwu Liu\IEEEauthorrefmark{1},
Yinghui Ye\IEEEauthorrefmark{2},
Zhiquan Bai\IEEEauthorrefmark{3},
Kyeong Jin Kim\IEEEauthorrefmark{4},
and Theodoros A. Tsiftsis\IEEEauthorrefmark{5}}

\IEEEauthorblockA{
\IEEEauthorrefmark{1} School of Information Science and Electrical Engineering, Shandong Jiaotong University, Jinan 250357, China\\
\IEEEauthorrefmark{2} Shaanxi Key Laboratory of Information Communication Network and Security, Xi'an University of Posts \& Telecommunications\\
\IEEEauthorrefmark{3} School of Information Science and Engineering, Shandong University, Qingdao 266237, China\\
\IEEEauthorrefmark{4} Mitsubishi Electric Research Laboratories, Cambridge, MA 02139, USA\\
\IEEEauthorrefmark{5} School of Intelligent Systems Science and Engineering, Jinan University, Zhuhai 519070, China\\
Email:~liuhongwu@sdjtu.edu.ch, connectyyh@126.com, zqbai@sdu.edu.cn, kkim@merl.com, theo\_tsiftsis@jnu.edu.cn}
}

\maketitle

\setcounter{page}{1}
\begin{abstract}
In this paper, we investigate rate splitting multiple access (RSMA) aided mobile edge computing (MEC) in a cognitive radio network. We propose a RSMA scheme that enables the secondary user to offload  tasks to the MEC server utilizing dynamic rate splitting without deteriorating the primary user's offloading. The expressions for the optimal rate splitting parameters that maximize the achievable rate for the secondary user and successful computation probability of the proposed RSMA scheme are derived in closed-form. We formulate a problem to maximize  successful computation probability by jointly optimizing  task offloading ratio and task offloading time and obtain the optimal solutions in closed-form. Simulation results clarify that the proposed RSMA scheme achieves a higher successful computation probability than the existing non-orthogonal multiple access scheme.
\end{abstract}


\section{Introduction}

For the forthcoming computation-intensive and delay-sensitive services, e.g., virtual reality, smart manufacturing, and autonomous driving, mobile edge computing (MEC) has been deemed as a promising technology to cope with  the challenges of providing superior and latency-critical computing \cite{MEC_Survey}. Equipped with powerful computing and storage capabilities, MEC servers enable edge users to offload their tasks for nearby processing. Thus, backhaul bottlenecks, network delays, and transmission costs can be significantly reduced.

Driven by the benefits of MEC, massive edge users may need to offload tasks to MEC servers for computing, which greatly intensifies spectrum scarcity.
As an essential role for dynamic spectrum access, cognitive radio (CR) technology has been widely utilized to improve spectrum efficiency (SE) \cite{CR:1}. In CR networks, secondary users (SUs) can access spectrum bands licensed to primary users (PUs) by controlling access parameters subject to interference temperature constraints. Recently, studies have been found in literatures for improving energy efficiency (EE) and SE and decreasing offloading latency in CR-based MEC (CR-MEC) networks \cite{CR_MEC_industrial,Distributed_CR_MEC_NOMA,Delay_NOMA_CR_MEC, Offloading_CR_NOMA_MEC}. Specifically, non-orthogonal multiple access (NOMA) technology has been widely investigated to increase SE/EE and reduce latency for CR-MEC networks \cite{Distributed_CR_MEC_NOMA,Delay_NOMA_CR_MEC, Offloading_CR_NOMA_MEC}. Even without the implicit interference temperature constraint, the paired users in NOMA-aided MEC networks can be regarded as that the PU and SU cooperatively to offload tasks to reduce latency \cite{NOMA_MEC_Psucc}. Nevertheless, the decoding order of successive interference cancellation (SIC) strongly affects the system performance of CR-inspired NOMA (CR-NOMA) \cite{Unveiling_SIC_PartI}, so that the hybrid SIC has been utilized to minimize the offloading energy for CR-NOMA enabled MEC \cite{Unveiling_SIC_PartII}.

It is well known that uplink NOMA with the SIC receiver can achieve only several separated points on the rate boundary of multiple access channels, whereas uplink rate splitting multiple access (RSMA) can achieve the full rate boundary \cite{Rate_Splitting_MAC}. Recently, RSMA technique has been investigated for various wireless communications scenarios (see \cite{RSMA_Trends} and references therein). Motivated by this, we have proposed a CR-inspired rate splitting scheme to enhance the outage performance and user fairness for uplink NOMA systems \cite{RS_NOMA_UL_fair}. In this paper, we propose a new RSMA scheme to aid MEC for a CR network. The proposed RSMA scheme allows the SU to simultaneously offload with the PU through dynamic rate splitting, meanwhile avoiding the deterioration of the PU's offloading. As part of the proposed RSMA scheme, we firstly derive the closed-form expressions for the optimal rate splitting parameters that attain the maximum achievable rate for the SU. Then, a problem is formulated to maximize the successful computation probability by optimizing the task offloading ratio and task offloading time, whose optimal solutions are also derived in closed-form. Various simulation results are provided to clarify the superior performance of the proposed RSMA scheme.

\section{System Model}

The considered CR network consists of one MEC server, a PU $U_a$, and a SU $U_b$. Each node is equipped with a single antenna for transmitting or receiving. The channel from $U_k$ ($k \in \{a, b \}$) to the MEC server is denoted by $\ell_k h_k$, where $h_k$ is the small-scale fading coefficient and $\ell_k = (1+d_k^\upsilon)^{-1}$ is the path-loss with the distance from $U_k$ to the MEC server, $d_k$, and the path-loss exponent, $\upsilon$. We assume the quasi-static fading channels, i.e., the channel fading coefficients keep constant within one transmission block and may vary from one transmission block to another one.
The small-scale channel fading coefficients $h_a$ and $h_b$ are modeled as independent and identically distributed (i.i.d.) circular symmetric complex Gaussian random variables with zero mean and unit variance.

In the considered CR network, $U_k$ has the computation tasks with $M_k$ bits, which are bit-wise independent and can be divided into different groups arbitrarily. Due to the limited local computation capability, $U_k$ may not be able to execute all the tasks locally within the latency budget. Thus, $U_k$ offloads its tasks partially with $\eta_k M_k$ bits to the MEC server and executes the remaining tasks with $(1-\eta_k) M_k$ bits locally, where $\eta_k$ ($0 \le \eta_k \le 1$) is the task offloading ratio. To facilitate the task offload within the latency budget of $T$ seconds, which is less than the coherence time of channel fading, two users simultaneously transmit their task-bearing signals to the MEC server using RSMA. With respect to the limited execution capabilities of the users, we assume that each user can operate in either transmission or computation and consider a hybrid offloading scheme that supports three modes, namely, partial computation offloading, full local computation, and complete offloading.

The working flow of the considered offloading scheme is described as follows. In the first phase $t_1$, the MEC server determines the offloading and rate splitting parameters. Then, the PU and SU offload their tasks to the MEC server using RSMA in the second phase $t_2$. In the third phase $t_3$, the successfully offloaded tasks are computed at the MEC server, meanwhile the PU and SU compute their remaining tasks locally. In the last phase $t_4$, the MEC server feeds back the computed results of the offloaded tasks to the PU and SU, respectively.
Considering that $t_1$ and $t_4$ are very small compared to $t_2$ and $t_3$, we ignore $t_1$ and $t_4$ and only consider $t_2$ and $t_3$ in the offloading scheme, i.e., the offloading scheme consists of the offloading phase $t_2$ and the task computing phase $t_3$ under the constraint $t_2 + t_3 \le T$.

{{\emph{A. Offloading Phase:}}} In the offloading phase, the PU and SU transmit their signals bearing $\eta_a M_a$ and $\eta_b M_b$ task bits to the MEC server, respectively. Let $x_k$ denote the signal of $U_k$, the SU  splits its signal $x_b$ into two streams, $x_{b,1}$ and $x_{b,2}$ using the rate splitting parameters determined in the phase $t_1$. Then, the received signal at the MEC server can be expressed as follows:
\begin{eqnarray}
    y &\!\!\!  =  \!\!\! &  \sqrt{P_a \ell_a} h_a x_a +  \sqrt{\alpha P_b  \ell_b} h_b x_{b, 1}   \nonumber \\
    &\!\!\!   \!\!\! &  +  \sqrt{(1 -  \alpha) P_b \ell_b} h_b x_{b, 2}  +  w,~~~~ \label{eq:y}
\end{eqnarray}
where $P_k$ is the transmit power of $U_k$, $\alpha$ is the power allocation factor satisfying $0 \le \alpha \le 1$, and $w$ is the additive white Gaussian noise at the MEC server with zero mean and variance $\sigma^2$. We assume that all the signals, i.e., $\{x_a, x_{b,1}, x_{b,2}\}$, are independently coded using Gaussian code book with zero mean and unit variance.

At the MEC server, the decoding order $x_{b,1} \to x_a \to x_{b,2}$ is utilized for the SIC processing with the aim to obtain the maximum allowed achievable rate for the SU. The received signal-to-noise-plus-interference-ratios (SINRs) and signal-to-noise-ratio (SNR) for decoding $x_{b,1}$, $x_a$, and $x_{b,2}$ can be expressed as:
\begin{eqnarray}
   \gamma_{b,1} = \frac{\alpha  \rho_b  |h_b|^2}{ \rho_a   |h_a|^2 + (1-\alpha)\rho_b  |h_b|^2 + 1},   \label{eq:snr_rb1}
\end{eqnarray}
\begin{eqnarray}
    \gamma_a =  \frac{\rho_a  |h_a|^2}{(1-\alpha) \rho_b  |h_b|^2 + 1} , \label{eq:snr_ra}
\end{eqnarray}
\begin{eqnarray}
    \gamma_{b,2} =  (1-\alpha)\rho_b  |h_b|^2,  \label{eq:snr_rb2}
\end{eqnarray}
respectively, where $\rho_a \triangleq \frac{P_a \ell_a}{\sigma^2}$ and $\rho_b \triangleq \frac{P_b \ell_b}{\sigma^2}$ denote the equivalent transmit SNRs of $U_a$ and $U_b$, respectively. Let $B$ denote the allocated bandwidth for the two users, the achievable rates for transmitting $x_{a}$ and  $x_{b}$ can be written as:
\begin{eqnarray}
R_{a} = t_2 B \log_2(1 +\gamma_{a}) {~~{\text{and}}~~}  R_{b} = R_{b,1} + R_{b,2},
\end{eqnarray}
respectively, with $R_{b,1} = t_2 B \log_2(1 +\gamma_{b,1})$ and $R_{b,2} = t_2 B \log_2(1 +\gamma_{b,2})$.

For the transmissions of $x_a$, $x_{b,1}$, and $x_{b,2}$, the target rates $\widehat R_a \triangleq \eta_a M_a$, $\widehat R_{b,1} \triangleq \beta \eta_b M_b$, and $\widehat R_{b,2} \triangleq (1-\beta) \eta_b M_b$ are defined, respectively, where $\beta$ is the rate splitting factor satisfying $0 \le \beta \le 1$.
If $x_{a}$, $x_{b,1}$, and $x_{b,2}$ are decoded successfully, the total offloaded tasks can be expressed as:
\begin{eqnarray}
    M_{_{\rm MEC}} &\!\!\!=\!\!\!&  \widehat R_a +  \widehat R_{b,1} + \widehat R_{b,2} =  \eta_a  M_a +  \eta_{b} M_b.
\end{eqnarray}

{{\emph{B. Task Executing Phase:}}} In the task executing phase, the offloaded tasks $M_{_{\rm MEC}}$ is executed at the MEC server, meanwhile the local tasks $(1-\eta_a)M_a$ and $(1-\eta_b)M_b$ are executed at $U_a$ and $U_b$, respectively. The execution time of the offloaded and local tasks can be respectively computed by
\begin{eqnarray}
    t_3^{^{\rm MEC}} = \frac{M_{_{\rm MEC}} C}{f_{_{\rm MEC}}} {~~{\text{and}}~~}  t_3^{^{U_k}} = \frac{(1-\eta_k)M_k C}{f_{\rm user}},  \label{eq:t2_user}
\end{eqnarray}
where $f_{_{\rm MEC}}$ and $f_{\rm user}$ denote the CPU frequencies at the MEC server and the two users, respectively, and $C$ is the number of the CPU cycles required for computing one task bit.
Considering that the MEC server has a much stronger computation capability over the two users, we assume that $f_{_{\rm MEC}} = N f_{\rm user}$ with $N > 1$.

\section{RSMA Scheme}

In the considered CR network, the SU is allowed to share the same time/frequency resource occupied by the PU without introducing intolerable interference. In the proposed RSMA scheme, the task offloading from the SU to the MEC server is permitted only when the task offloading performance of the PU is guaranteed. In other words, the proposed RSMA scheme ensures that the delay-limited task offloading of the PU does not deteriorate compared to the scenario where the PU occupies the time/frequency resource alone. In this section, the design principle, operations, and optimal rate splitting parameters of the proposed RSMA scheme are presented in details.

The RSMA scheme is designed to obtain the maximum allowed achievable rate for the SU, meanwhile keeping the PU to attain the same successful computation probability as that when the SU does not offload.
Considering that the decoding order $x_{b,1} \to x_a \to x_{b,2}$ is utilized at the MEC server for the SIC processing, the correct/incorrect detection of $x_{b,1}$ will affect the detections of both $x_{a}$ and $x_{b,2}$. Without loss of generality, we assume that $x_{b,1}$ has been correctly detected during the first stage of the SIC processing. Then, the remaining SIC processing is to sequentially detect $x_a$ and $x_{b,2}$ using the decoding order $x_a \to x_{b,2}$.

To ensure that the PU's offloading does not deteriorate during the SIC processing for $x_a$ and $x_{b,2}$, an interference threshold $\tau$ is computed by the MEC server as:
\begin{eqnarray}
    \tau  = \max\left\{0, \frac{\rho_a |h_a|^2}{\varepsilon_a}  -1  \right\}, \label{eq:threshold}
\end{eqnarray}
where $ \varepsilon_{a} \triangleq 2^{\tfrac{\eta_{a} M_a}{t_2 B}} - 1$.
With respect to the received SINR $\gamma_{a}$ as in \eqref{eq:snr_ra}, when the equivalent interference power associated with the transmission of $x_{b,2}$ is no more than $\tau$, i.e., $ (1-\alpha) \rho_b |h_b|^2 \le \tau$, we have the following relationships:
\begin{eqnarray}
\!\!\!\!\!\! \Pr \left( R_a \!\ge\! \eta_a M_a  \right)  &\!\!\!\!\ge\!\!\!\!& \Pr\left( t_2 B \log_2\left(1 \!+\! \frac{\rho_a|h_a|^2 }{ \tau + 1}  \right) \!\ge\! \eta_a M_a  \right)   \nonumber \\
\!\!\!\!\!\! &\!\!\!\! = \!\!\!\!&  \Pr\left( t_2 B \log_2\left(1 + \rho_a|h_a|^2 \right) \!\ge\! \eta_a M_a \right)\!.
\end{eqnarray}
As a result, in the presence of the SU's interference signal $\sqrt{(1-\alpha)P_b \ell_b} h_b x_{b,2}$ subject to $ (1-\alpha) \rho_b |h_b|^2 \le \tau$, the PU can successfully offload its task bits $\eta M_a$ to the MEC server with the same probability as that it occupies the time/frequency resource alone during the phase $t_2$.

Prior to transmission of the phase $t_2$, the MEC server sends the interference threshold $\tau$ to the SU. Then, the SU  compares its CSI with $\tau$ and determines its rate splitting operation and the optimal rate splitting parameters. Depending on the SU's  CSI and $\tau$, the rate splitting operations and the optimal rate splitting parameters are designed as follows:

{\emph{Case I:}} $0< \rho_b |h_b|^2 \le \tau$. In this case, the PU can successfully offload its task bits to the MEC server due to the fact that $\tau > 0$.
Since the detection of $x_{b,2}$ is interference-free at the last stage of the SIC processing, to maximize $R_b = R_{b,1} + R_{b,2}$ is equivalently to maximize $R_{b,2}$ in this case. Therefore, the SU allocates all of $P_b$ to transmit $x_{b,2}$ such that $R_{b,2} = t_2 B\log_2(1+\gamma_{b,2})$ is maximized.
On the contrary, zero power is allocated to transmit $x_{b,1}$ in this case.

As a result, the RSMA scheme sets $x_{b,2} = x_b$ in Case I and the optimal rate splitting parameters are given by
\begin{eqnarray}
\alpha^* = 0  {\rm~~and~~}  \beta^* = 0,
\end{eqnarray}
respectively, where $(\cdot)^*$ denotes the optimal solution for the corresponding parameter. Considering that  $x_{b,1}$ is not transmitted, the decoding order $x_{b,1} \to x_a \to x_{b,2}$ degrades to $x_a \to x_{b}$. Then, the achievable rates of $U_a$ and $U_b$ in Case I can be expressed as
\begin{eqnarray}
R_a^{(\rm I)} = t_2 B \log_2\left(1 + \frac{\rho_a |h_a|^2}{\rho_b |h_b|^2 + 1}  \right),
\end{eqnarray}
\begin{eqnarray}
R_b^{(\rm I)} = t_2 B \log_2\left(1 + \rho_b |h_b|^2  \right),
\end{eqnarray}
respectively.

\emph{Case II:} $0 < \tau < \rho_b |h_b|^2$. In this case, to maximize the achievable rate $R_b = R_{b,1} + R_{b,2}$, it needs to maximize $R_{b, 2} = t_2 B\log_2(1+\gamma_{b,2})$ firstly subject to the constraint of $R_{b, 2} \le  t_2 B\log_2(1+\tau)$ and the maximum $R_{b, 2}$ is achieved by setting $\gamma_{b, 2} = \tau$, which results in the maximum $R_{b, 2} = t_2 B\log_2(1+\tau)$ and the optimal power allocation factor
\begin{eqnarray}
\alpha^* = 1 -  \frac{1}{\rho_b |h_b|^2}\left( \frac{\rho_a |h_a|^2}{\varepsilon_a}  -1 \right).
\end{eqnarray}
Furthermore, the transmit power $\alpha P_b = \left( 1 -  \tfrac{\tau}{\rho_b |h_b|^2}  \right) P_b$ is allocated to transmit  $x_{b,1}$, which results in the maximum achievable rate $R_{b,1} = t_2 B\log_2\left( 1 + \tfrac{\rho_b |h_b|^2 - \tau}{\rho_b |h_b|^2 + \tau + 1} \right) $ in this case.
Consequently, the achievable rates of $U_a$ and $U_b$ in Case II are given by
\begin{eqnarray}
R_a^{({\rm II})}  = t_2 B \log_2\left( 1 + \frac{\rho_a |h_a|^2}{\tau + 1} \right),
\end{eqnarray}
\vspace{-0.1in}
\begin{eqnarray}
R_b^{({\rm II})} \!\!&\!\!\!\!=\!\!\!\!&\! t_2 B \log_2\left( 1 \!+\! \frac{\rho_b |h_b|^2 - \tau}{\rho_a |h_a|^2 + \tau + 1} \right) \!+\!  t_2 B \log_2(1 \!+\! \tau), \nonumber \\
\end{eqnarray}
respectively.

To avoid the error propagation in the SIC processing due to the failure detection of $x_{b,1}$, it is required that $R_b^{({\rm II})}  \ge \eta_b M_b $ to ensure the correct detection of $x_{b,1}$. In the proposed RSMA scheme, we set the target rate as $\widehat R_{b,2} = t_2 B \log_2(1  +  \tau)$, which is achieved by the transmission of $x_{b,2}$. Considering that $ \widehat R_{b,2} = (1-\beta) \eta_b M_b  $, the optimal rate splitting factor is determined by
\begin{eqnarray}
\beta^* = 1 - \frac{t_2 B \left( \log_2(\rho_a|h_a|^2)-\log_2(\varepsilon_2)\right)}{\eta_b M_b}.
\end{eqnarray}
As a result, only when $R_{b,1} \ge \widehat R_{b,1}$, where $\widehat R_{b,1} =  \eta_b M_b - \widehat R_{b, 2}$, the SU is permitted to transmit the signal using the rate splitting; Otherwise, the SU keeps silence in Case II. Nevertheless, the PU can attain the achievable rate $R_a^{(\rm I)} = t_2 B \log_2(1 + \rho_a |h_a|^2)$ even if the SU keeps silence.

\emph{Case III:} $\tau = 0$. When $\tau = 0$, the PU cannot successful offload its task bits $\eta_a M_a$ to the MEC server. To utilize the transmit power efficiently, the SU allocates all of $P_b$ to transmit $x_{b,1} = x_b$ to ignore the failure detection of $x_a$. Consequently, the optimal rate splitting parameters are given by $\alpha^* =1$ and $\beta^* = 1$, respectively. Then, the SIC decoding order $x_{b,1} \to x_a \to x_{b,2}$ degrades to $x_b \to x_a$ and the achievable rates are given by
\begin{eqnarray}
R_a^{(\rm III)} = t_2 B \log_2(1 + \rho_a |h_a|^2),
\end{eqnarray}
\begin{eqnarray}
R_b^{(\rm III)} = t_2 B \log_2\left(1 + \frac{\rho_b |h_b|^2}{\rho_a |h_a|^2 + 1} \right),
\end{eqnarray}
respectively.

{\bf\emph{Remark 1:}} In Case I, the operation of the proposed RSMA scheme is the same as that of the two-user NOMA system utilizing the SIC decoding order $x_a \to x_b$ such that the maximum achievable rate of $U_b$ is attained. In Case II, $U_b$ dynamically adjusts its rate splitting parameters to maximize its achievable rate while the PU still can offload its task bits $\eta_a M_a$ to the MEC server successfully as in OMA with delay-limited transmissions. Moreover, in Case III, the operation of the proposed RSMA scheme is the same as that of the two-user NOMA system utilizing the SIC decoding order $x_b \to x_a$, which also yields the maximum achievable rate for $U_b$.

\section{Successful Computation Probability Maximization}

To evaluate the MEC performance of the considered CR network, we first derive the closed-from expression for the successful computation probability achieved by the proposed RSMA scheme. Then, we determine the optimal offloading parameters that can achieve the maximum successful computation probability.

The successful computation probability is defined as the probability that all the offloaded and local tasks are successfully executed within the task executing phase \cite{NOMA_MEC_Psucc}. For the considered CR network, the successful computation probability achieved by the proposed RSMA scheme can be written as
\begin{eqnarray}
{\cal P}_s = {\cal P}_s^{(\rm I)} + {\cal P}_s^{(\rm II)} + {\cal P}_s^{(\rm III)},
\end{eqnarray}
where
\begin{eqnarray}
    {\cal P}_s^{(\rm I)} &\!\!\!=\!\!\!& \Pr\left\{ 0< \rho_b|h_b|^2  \le  \tau, R_a^{\rm(I)}  \ge  \eta_a M, R_b^{\rm(I)}  \ge  \eta_b M, \right.  \nonumber \\
     &\!\!\! \!\!\!&  ~~~~  \left. \max\left(t_3^{^{U_a}}, t_3^{^{U_b}}, t_3^{^{\rm MEC}} \right) \le t_3 \right\},
\end{eqnarray}
\begin{eqnarray}
    {\cal P}_s^{(\rm II)} &\!\!\!=\!\!\!& \Pr\left\{ 0< \tau < \rho_b|h_b|^2, R_a^{\rm(II)}  \ge  \eta_a M, R_b^{\rm(II)}  \ge  \eta_b M , \right.  \nonumber \\
     &\!\!\! \!\!\!&  ~~~~  \left. \max\left(t_3^{^{U_a}}, t_3^{^{U_b}}, t_3^{^{\rm MEC}} \right) \le t_3  \right\},
\end{eqnarray}
\begin{eqnarray}
    {\cal P}_s^{(\rm III)}  &\!\!\!=\!\!\!& \Pr\left\{ \tau \!=\!0, R_a^{\rm(III)}  \ge  \eta_a M, R_b^{\rm(III)}  \ge \eta_b M , \right.  \nonumber \\
     &\!\!\! \!\!\!&  ~~~~  \left. \max\left(t_3^{^{U_a}}, t_3^{^{U_b}}, t_3^{^{\rm MEC}} \right) \le t_3  \right\}
\end{eqnarray}
correspond to the successful computation probability achieved by the RSMA scheme in Cases I, II, and III, respectively.
Taking into account that the successful computation probability is 0 when the event $\max\left(t_3^{^{U_a}}, t_3^{^{U_b}}, t_3^{^{\rm MEC}} \right) > t_3$ occurs \cite{NOMA_MEC_Psucc},
hereafter we only consider the successful computation probability in the case of $\Pr\left\{ \max\left(t_3^{^{U_a}}, t_3^{^{U_b}}, t_3^{^{\rm MEC}}  \right) \le t_3 \right\} = 1$.

{\bf{Theorem 1:}}
    Closed-from expression for the successful computation probability achieved by the RSMA scheme is given by
    \begin{eqnarray}
         {\cal P}_s = \frac{\rho_a e^{-\tfrac{\varepsilon_b}{\rho_b} - \tfrac{\varepsilon_a(1+\varepsilon_b)}{\rho_a}} }{\rho_a - \rho_b} - \frac{\rho_b e^{-\tfrac{\varepsilon_a}{\rho_a} - \tfrac{\varepsilon_b(1+\varepsilon_a)}{\rho_b}} }{\rho_a - \rho_b},  \label{eq:Psucc_CR}
    \end{eqnarray}
    where $ \varepsilon_{b} \triangleq 2^{\frac{\eta_{b} M_b}{t_2 B}} - 1$.

\begin{IEEEproof}
    Please see Appendix A.
\end{IEEEproof}

{\bf\emph{Remark 2:}}  Proposition 1 provides the closed-form expression for the successful computation probability in terms of statistic CSI and offloading parameters, which enables an easy way for the computer evaluation of the successful computation probability. The results in Proposition 1 also offer a possibility to obtain the optimal parameters that maximize the successful computation probability without requiring the instantaneous CSI, which is preferred for alleviating the signalling burdens.

{\bf\emph{Remark 3:}}  When $\frac{M C}{f_{\rm user}} \le T$ occurs, the successful computation probability ${\cal P}_s = 1$ always holds since that the PU and SU can successfully execute all the tasks locally within $T$. Consequently, the optimal offloading parameters are determined as follows: $\eta_{a}^*  = \eta_b^* = 0$, $t_2^* = 0$ and $t_3^* = \frac{M C}{f_{\rm user}}$. In such a case, since the rate splitting does not occur, it is unnecessary to optimize $\alpha$ and $\beta$.

When $\frac{M C}{f_{\rm user}} > T$, in order to achieve the allowed maximum successful computation probability, we formulate a successful computation probability maximization problem by jointly optimizing the offloading parameters, including $t_2$, $t_3$, $\eta_{a}$, and $\eta_b$, as follows:
\begin{eqnarray}
    {\bf{P}}_0: &\!\!\! \!\!\!& \max\limits_{t_2, t_3, \eta_{a}, \eta_b} {\cal{P}}_s  \nonumber \\
    &\!\!\! \!\!\!&\!\!\!  C_1: ~~ \frac{M C}{f_{\rm user}} > T, \nonumber \\
    &\!\!\! \!\!\!&\!\!\!  C_2: ~~ t_2>0, t_3>0, t_2 + t_3 \le T,  \label{eq:P0}  \\
    &\!\!\! \!\!\!&\!\!\!  C_3: ~~ 0 < \eta_{a}, \eta_b < 1.  \nonumber
\end{eqnarray}

Since the objective function in \eqref{eq:P0} is non-convex, the successful computation probability maximization problem $ {\bf{P}}_0$ is not a convex problem. On the optimal offloading parameters, we have the following proposition.

{\bf{Proposition 1:}}
    The maximum successful computation probability is achievable by satisfying the equalities $t_2 + t_3 = T$ and $t_3^{^{U_a}} = t_3^{^{U_b}} = t_3^{^{\rm MEC}} = t_3$.

\begin{IEEEproof}
Using the monotonicity of ${\cal P}_s$ with respect to $t_2$, $\eta_a$, and $\eta_b$, the results in Proposition 1 can be obtained via proof by contradiction.
\end{IEEEproof}

{\bf\emph{Remark 4:}} The intuition behind Proposition 1 is that all the computation capabilities of the MEC server and two users should be fully exploited within the time phase $t_3$ to execute the offloaded and local tasks, respectively, which results in the shortest duration for $t_2$ and guarantees that the maximum computation probability is achievable.


{\bf{Theorem 2:}}
    To achieve the maximum successful computation probability, the optimal offloading parameters are given by
    \begin{eqnarray}
        \eta_a^* =  \frac{(N+1)M_a - M_b}{(N + 2)M_a},  \label{eq:etaa*}
    \end{eqnarray}
    \vspace{-0.1in}
    \begin{eqnarray}
        \eta_b^* =  \frac{(N+1)M_b - M_a}{(N + 2)M_b},  \label{eq:etab*}
    \end{eqnarray}
    \vspace{-0.1in}
    \begin{eqnarray}
        t_2^*  = T - \frac{(M_a + M_b) C}{(N+2)f_{\rm user}},  \label{eq:t2*}
    \end{eqnarray}
    \vspace{-0.1in}
    \begin{eqnarray}
        t_3^*  = \frac{(M_a + M_b) C}{(N+2)f_{\rm user}},  \label{eq:t3*}
    \end{eqnarray}
    respectively.

\begin{IEEEproof}
    Substituting $t_3^{^{U_a}} = t_3^{^{U_b}} = t_3^{^{\rm MEC}} = t_3$, which is the constraint for achieving the maximum successful computation probability, into \eqref{eq:t2_user} and $t_2 + t_3 = T$, we immediately obtain the expressions \eqref{eq:etaa*}, \eqref{eq:etab*}, \eqref{eq:t2*}, and \eqref{eq:t3*} by solving the corresponding functions.
\end{IEEEproof}

{\bf\emph{Remark 5:}} The results in Theorem 2 characterize the optimal offloading parameters in terms of the computation and latency parameters, which makes it convenient for evaluating the maximum successful computation probability.
The expressions \eqref{eq:etaa*} and \eqref{eq:etab*} reveal  the impact of $N$ on the offloading ratio $\eta_a$ and $\eta_b$, respectively. Considering that $0<\eta_a^*<1$ and $0<\eta_b^*<1$, we have the following hide constraint $\frac{1}{N} < \frac{M_a}{M_b} < N+1$ on the task bits $M_a$ and $M_b$, i.e., the task bits of the two users should be comparable. Additionally, the expression \eqref{eq:t2*} results in the hide constraint $M_a + M_b < \frac{(N+2)f_{\rm user}T}{C}$, which indicates the maximum task bits that can be processed by the MEC server and two users.

{\bf\emph{Remark 6:}}
As $N \to \infty$, we have $ \eta_a^* \to 1$ and  $\eta_b^* \to 1$, i.e., the PU and SU prefer the complete offloading. If $N$ is a finite number, the PU and SU choose to  offload  the partial tasks to the MEC server. By fixing the offloading time phase $t_2$, the executing phase $t_3$ decreases with the increasing of $f_{\rm user}$ and $N$. Thus, the latency budget $T$ can be reduced.
As $\frac{P_a}{\sigma^2} \to \infty$ and $ \frac{P_b}{\sigma^2} \to \infty$, asymptotic expressions for ${\cal P}_s^{(\rm I)}$ and ${\cal P}_s^{(\rm II)}$ in the high SNR regime can be expressed as:
\begin{eqnarray}
    {\cal P}_s^{(\rm I)} = \frac{\ell_a}{\ell_b \varepsilon_a + \ell_a } {~~{\text{and}}~~}  {\cal P}_s^{(\rm II)}   = \frac{\ell_b \varepsilon_a}{\ell_b \varepsilon_a + \ell_a \gamma_a},
\end{eqnarray}
respectively, which yields ${\cal P}_s^{(\rm I)} + {\cal P}_s^{(\rm II)} = 1$ in the high SNR region.
Recalling that ${\cal P}_s^{(\rm III)}=0$ in Case III, we have ${\cal P}_s =1$ as $\frac{P_a}{\sigma^2} \to \infty$ and $ \frac{P_b}{\sigma^2} \to \infty$.

\begin{figure}[tb]
    \begin{center}
    \includegraphics[width=3.2in]{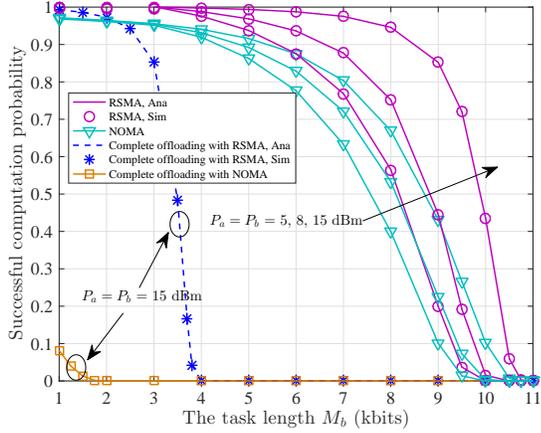}
    \vspace{-0.1in}
    \caption{${\cal P}_s$ versus the task length.}
    \label{fig:subfig3c}
    \end{center}
    \vspace{-0.15in}
\end{figure}

\begin{figure}[tb]
    \begin{center}
    \includegraphics[width=3.2in]{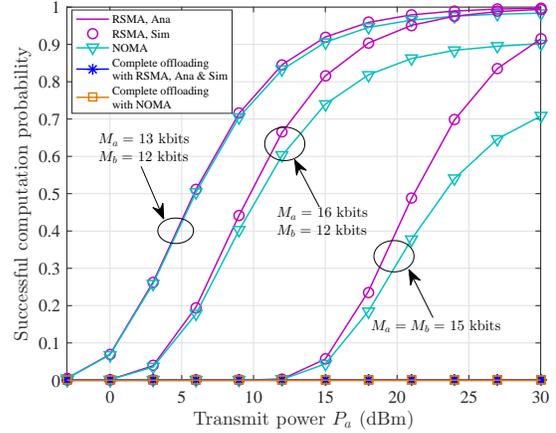}
    \vspace{-0.1in}
    \caption{${\cal P}_s$ versus the transmit power ($P_a = P_b$).}
    \label{fig:subfig3c}
    \end{center}
    \vspace{-0.15in}
\end{figure}

\begin{figure}[tb]
    \begin{center}
    \includegraphics[width=3.2in]{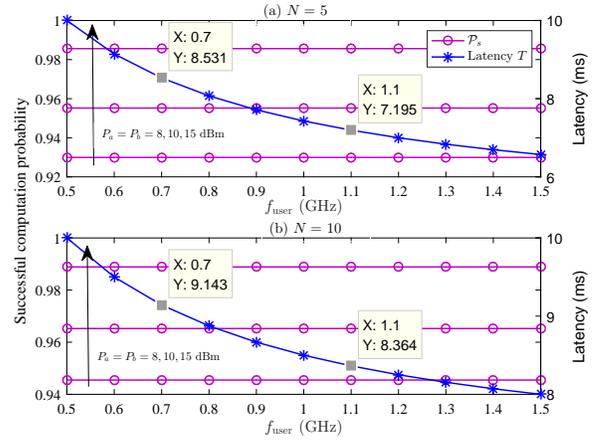}
    \vspace{-0.1in}
    \caption{The way to reduce latency.}
    \label{fig:subfig3c}
    \end{center}
    \vspace{-0.15in}
\end{figure}

\section{Simulation Results}

In this section, we present the simulation results to clarify the system performance achieved by the proposed RSMA scheme and verify the accuracy of the developed analytical expressions. Unless otherwise specified, the following parameters are used throughout the simulation: $B=1$ MHz, $C=1000$ cycle/bit, $N = 5$, $T = 10$ ms, $d_a=5$ m, $d_b=25$ m, $\upsilon=4$, $f_{\rm user} = 0.5$ GHz, and $\sigma^2=10^{-9}$ W.

In Fig. 1, the successful computation probability achieved by the proposed RSMA scheme is compared with the NOMA scheme of \cite{NOMA_MEC_Psucc} for various task length settings. Particularly, we assume $M_a = 20$ kbits and varies $M_b$ from 1 kbits to 11 kbits in Fig. 1.
The curves in Fig. 1 verify the accuracy of the analytical result in Theorem 1. With the increasing of task length $M$, the successful computation probability achieved by the proposed RSMA scheme decreases due to the decreasing of $t_2$. We can see that the proposed RSMA scheme achieves the higher successful computation probability over that of the NOMA scheme. Moreover, the successful computation probability achieved by the complete offloading using the proposed RSMA is higher than that of the complete offloading scheme using NOMA. For the considered task length settings, the full local computing scheme always achieves ${\cal P}_s = 0$, so that we omit it in Fig. 1. The results in Fig. 1 clarify that the proposed RSMA scheme outperforms the NOMA scheme in the middle and high  task length regions.

In Fig. 2, the curves of the successful computation probability versus various transmit power settings are presented. The curves in Fig. 2 also verify the correctness of the analytical expression for the successful computation probability achieved by the proposed RSMA scheme. It can be seen from Fig. 2 that the successful computation probability achieved by the proposed RSMA scheme increases with the increasing of the transmit power. The reason for this phenomenon is that a larger transmit power brings a larger $\rho_a$ ($\rho_b$), which results in a larger ${\cal P}_s$. Moreover, the RSMA scheme achieves a higher successful computation probability than that of the NOMA scheme in the considered middle and high transmit power regions.

In Fig. 3, we presents the numerical results to show the way to reduce latency as discussed in \emph{Remark 6}. The simulation parameters are set as $M_a = 10$ kbits, $M_b = 8$ kbits, and $P_a = P_b$.  By fixing $t_2 =  10 - \frac{(M_a + M_b)C}{(N+2)f_{\rm user}}$, while $t_3$ is computed by \eqref{eq:t3*}, the way of reducing $T = t_2 + t_3$ is found. From Fig. 3(a), where the left vertical axis indicates ${\cal P}_s$ and the right vertical axis indicates the latency $T$, the latency $T$ decreases with the increasing of $f_{\rm user}$, while ${\cal P}_s$ remains unchanged. The reason for this phenomenon is that $t_2$ keeps unchanged while $t_3$ decreases due to the increasing of $f_{\rm user}$. The curves in Fig. 3(b) also show that ${\cal P}_s$ increases with the cost of higher latency due to a larger offloading ratio with the increasing of $N$.

\section{Conclusions}

We have proposed a new RSMA scheme to assist MEC in a CR network. During the offloading, the SU dynamically adjusts its rate splitting parameters without deteriorating the PU's offloading, such that the SU's achievable rate is maximized. We have derived closed-form expression for the successful computation probability achieved by the proposed RSMA scheme. The optimal offloading and rate splitting parameters have been obtained in closed-form. The superior performance of the proposed RSMA scheme over that of the existing NOMA scheme has been verified by various simulation results.

\section*{Appendix A: A proof of Theorem 1}

To derive closed-form expression for the successful computation probability, we evaluate ${\cal P}_s^{(\rm I)}$, ${\cal P}_s^{(\rm II)}$, and ${\cal P}_s^{(\rm III)}$, respectively.

Corresponding to Case I, ${\cal P}_s^{(\rm I)}$ can be rewritten as:
\begin{eqnarray}
    {\cal P}_s^{(\rm I)} &\!\!\!=\!\!\!& \Pr \left(|h_a|^2 > \frac{\varepsilon_a(1 + \rho_b |h_b|^2)}{\rho_a},  |h_b|^2 > \frac{\varepsilon_b}{\rho_b}  \right).~~
\end{eqnarray}
Considering that $|h_i|^2$ ($i = a, b$) follows exponential distribution, ${\cal P}_s^{(\rm I)}$ can be further evaluated as:
\begin{eqnarray}
  \! {\cal P}_s^{(\rm I)} &\!\!\!\!=\!\!\!\!&\!\!   \int\nolimits_{\tfrac{\varepsilon_b}{\rho_b} }^{\infty}  \!\!\int\nolimits_{\tfrac{\varepsilon_a(1 + \rho_b y)}{\rho_a}}^{\infty}   e^{-x} e^{-y}  dx dy =   \frac{\rho_a e^{ -\tfrac{\varepsilon_b}{\rho_b} - \tfrac{\varepsilon_a(1+\varepsilon_b)}{\rho_a} } }{\rho_b \varepsilon_a + \rho_a},~~~~~      \label{ap:PsuccI}
\end{eqnarray}
where $ \varepsilon_{b} \triangleq 2^{\frac{\eta_{b} M_b}{t_2 B}} - 1$.

Corresponding to Case II, we can evaluate ${\cal P}_s^{(\rm II)}$ as
\begin{eqnarray}
    {\cal P}_s^{(\rm II)} &\!\!\!=\!\!\!& \Pr \left(|h_a|^2 > \frac{\varepsilon_a}{\rho_a},  |h_b|^2 > \frac{\rho_a |h_a|^2 \varepsilon_a^{-1} -1}{\rho_b}, \right. \nonumber \\
    &\!\!\! \!\!\!& ~~~~~ \left. |h_b|^2 \ge \frac{ \varepsilon_a +\varepsilon_b + \varepsilon_a \varepsilon_b  -  \rho_a |h_a|^2  }{\rho_b}  \right)  \nonumber \\
    &\!\!\!=\!\!\!&  \Pr\left( |h_a|^2 > \frac{\varepsilon_a(1+\varepsilon_b)}{\rho_a}, |h_b|^2 >  \frac{\rho_a |h_a|^2 \varepsilon_a^{-1} -1}{\rho_b}   \right)  \nonumber \\
    &\!\!\! \!\!\!& + \Pr\left( \frac{\varepsilon_a}{\rho_a} < |h_a|^2 <  \frac{\varepsilon_a(1+\varepsilon_b)}{\rho_a}, \right. \nonumber \\
    &\!\!\! \!\!\!&  ~~~~~ \left. |h_b|^2 > \frac{ \varepsilon_a +\varepsilon_b + \varepsilon_a \varepsilon_b -  \rho_a |h_a|^2  }{\rho_b}  \right)   \nonumber \\
    &\!\!\!=\!\!\!&   \int\nolimits_{\tfrac{\varepsilon_a(1+\varepsilon_b)}{\rho_a} }^{\infty}  \int\nolimits_{\tfrac{\rho_a \varepsilon_a^{-1}y -1}{\rho_b}}^{\infty }   e^{-x} e^{-y}  dx dy \nonumber \\
    &\!\!\! \!\!\!& + \int\nolimits_{\frac{\varepsilon_a}{\rho_a} }^{\tfrac{\varepsilon_a(1+\varepsilon_b)}{\rho_a}}  \int\nolimits_{\tfrac{ \varepsilon_a +\varepsilon_b + \varepsilon_a \varepsilon_b -  \rho_a y  }{\rho_b}}^{\infty }   e^{-x} e^{-y}  dx dy  \nonumber \\
    &\!\!\!=\!\!\!&  \frac{\rho_b \left(e^{-\tfrac{\varepsilon_b}{\rho_b} - \tfrac{\varepsilon_a(1+\varepsilon_b)}{\rho_a}}  - e^{- \tfrac{\varepsilon_a}{\rho_a} -\tfrac{\varepsilon_b(1+\varepsilon_a)}{\rho_b} }  \right) }{\rho_a - \rho_b}  \nonumber \\
    &\!\!\! \!\!\!&   +  \frac{\rho_b \varepsilon_a e^{ -\tfrac{\varepsilon_b}{\rho_b} - \tfrac{\varepsilon_a(1+\varepsilon_b)}{\rho_a} } }{\rho_b \varepsilon_a + \rho_a}.      \label{ap:PsuccII}
\end{eqnarray}

\addtolength{\topmargin}{0.05in}

Since that $\tau = 0$ in Case III, we have $\Pr\big\{R_a^{(\rm III)}  \ge \eta_a M_a\big\} = 0$, which results in ${\cal P}_s^{(\rm III)} = 0$. By combining the derived ${\cal P}_s^{(\rm I)}$, ${\cal P}_s^{(\rm II)}$, and ${\cal P}_s^{(\rm III)}$, we readily arrive at \eqref{eq:Psucc_CR}.

\begin{balance}
\bibliography{IEEEabrv,IEEE_bib}
\end{balance}

\end{document}